\documentclass{esannV2}
\usepackage[latin1]{inputenc}
\usepackage{amssymb,amsmath,array}
\usepackage{url}
\usepackage{graphicx}
\usepackage{dsfont}
\usepackage{xcolor}
\usepackage{subfigure}

\graphicspath{{images/}}

\begin{document}
\title{Reducing offline evaluation bias of \\collaborative filtering algorithms}

\author{Arnaud de Myttenaere$^{1,2}$, Boris Golden$^1$, B\'en\'edicte Le
Grand$^3$ \& Fabrice Rossi$^2$
\vspace{.3cm}\\
1 - Viadeo\\
30 rue de la Victoire, 75009 Paris - France
\vspace{.1cm}\\
2 - Universit\'e Paris 1 Panth\'eon - Sorbonne - SAMM EA 4534 \\
90 rue de Tolbiac, 75013 Paris - France
\vspace{.1cm}\\
3 - Universit\'e Paris 1 Panth\'eon - Sorbonne - Centre de Recherche en
Informatique \\
90 rue de Tolbiac, 75013 Paris - France\\
}

\maketitle
\begin{abstract}
Recommendation systems have been integrated into the majority of large online
systems to filter and rank information according to user profiles. It thus
influences the way users interact with the system and, as a consequence,
bias the evaluation of the performance of a recommendation algorithm computed using
historical data (via \emph{offline evaluation}). This paper presents a new application of a weighted offline evaluation to reduce this bias for collaborative filtering  algorithms.
\end{abstract}

\section{Introduction}
Recommendation systems have been very frequently studied in the literature and aim to provide a user with a set of possibly ranked items that are supposed to match the interests of the user \cite{park2012literature}. Applications of such systems are ubiquitous in the Internet (e-commerce, online advertising, social networks, ...), and can be seen as a way to adapt a system to a user.

Obviously, recommendation algorithms must be evaluated before and during their active use in order to ensure their performance. Live monitoring is generally achieved using online performance metrics (e.g. click-through rate of displayed ads) whereas offline evaluation is computed using historical data. Offline evaluation allows to quickly test several strategies without having to wait for real metrics to be collected nor impacting the performance of the online system. One of the main strategies of offline evaluation consists in simulating a recommendation by removing a confirmation action (click, purchase, etc.)  from a user profile and testing whether the item associated to this action would have been recommended based on the rest of the profile \cite{shani2011evaluating}. 

As presented in \cite{li2011unbiased, demytt2014reducing} this scheme ignores various factors that have influenced historical data as the recommendation algorithms previously used, promotional offers on some specific products, etc. Even if limits of evaluation strategies for recommendation algorithms have been identified (\cite{HerlockerEtAl2004Evaluating,mcnee2006being,said2013user}), this protocol is still intensively used in practice. 

We study in this paper the general principle of instance weighting proposed in \cite{demytt2014reducing} and show its practical relevance beyond the simple case of constant recommendation (i.e. if recommendations are the same for every user). In addition to its good performances, this method is more realistic than solutions proposed in \cite{HerlockerEtAl2004Evaluating,mcnee2006being} for which a data collection phase based on random recommendations has to be performed. While this phase allows one to build a bias free evaluation data set, it has also adverse effects in terms of e.g. public image or business performance when used on a live system.

The rest of the paper is organized as follows. Section \ref{sec:problem-formulation} describes in details the setting and the problem. Section \ref{sec:reduc-eval-bias} introduces the weighting scheme proposed to reduce the evaluation bias. Section \ref{sec:exper-eval} demonstrates the practical relevance of our method on real world data extracted from Viadeo (professional social network\footnote{See \url{http://corporate.viadeo.com/en/} for more information about Viadeo.}).

\section{Problem formulation}\label{sec:problem-formulation}

\subsection{Notations and setting}
We denote $U$ the set of users, $I$ the set of items and $\mathcal{D}_t$ the historical data available at time $t$. A recommendation algorithm is a function $g$ from $U\times \mathcal{D}_t$ to some set built from $I$. We will denote $g_t(u) = g(u,\mathcal{D}_t)$ the recommendation computed by $g$ at instant $t$ for user $u$.
We assume given a quality function $l$ from the product of the result space of $g$ and $I$ to $\mathbb{R}^+$ that measures to what extent an item $i$ is correctly recommended by $g$ at time $t$ via $l(g_t(u),i)$. We denote $I_u$ the items associated to a user $u$.

Offline evaluation is based on the possibility of ``removing'' any item $i$ from a user profile. The result is denoted $u_{-i}$ and $g_t(u_{-i})$ is the recommendation obtained at instant $t$ when $i$ has been removed from the profile of user $u$.

Finally, offline evaluation follows a general scheme in which a user is chosen according to some probability on users $P(u)$, which might reflect the business importance of the users. Given a user, an item $i$ is chosen among the items associated to its profile, according to some conditional probability on items $P(i|u)$. When an item $i$ is not associated to a user $u$ (that is $i\not\in I_u$), $P(i|u)=0$. A very common choice for $P(u)$ is the uniform probability on $U$ and it is also very common to use a uniform probability for $P(i|u)$ (other strategy could favor items recently associated to a profile). As the system evolves over the time, $P(u)$ and $P(i)$ depends on $t$.

The two distributions $P(u)$ and $P(i|u)$ lead to a joint distribution $P(u,i)=P(i|u)P(u)$ on $U\times I$. 

\subsection{Origin of the bias in offline evaluation}
The classic offline evaluation procedure consists in calculating the quality of the recommendation algorithm $g$ at instant $t$ as
$L_t(g)=\mathbb{E}(l(g_t(u_{-i}),i))$ where the expectation is taken with respect
to the joint distribution:
\begin{equation}\label{eq:loss}
L_t(g)=\sum_{(u,i)\in U\times I}P_t(i|u)P_t(u)l(g_t(u_{-i}),i).
\end{equation}

Then if two algorithms are evaluated at two different moments, their qualities are not directly comparable. Although as in an online system $P(i|u)$ evolves over time\footnote{even if $P(u)$ could also evolve over time we do not consider the effects of such evolution in the present article.} once a recommendation algorithm is chosen based on a given state of the system, it starts influencing the state of the system when put in production, inducing an increasing distance between its evaluation environment (i.e. the initial state of the system) and the evolving state of the system. This influence is responsible for a bias on offline evaluation as it relies on historical data.

A naive solution to this bias would be to compare algorithms only with respect to the original database at $t_0$, but it would discard natural evolutions of user profiles. 

\section{Reducing the evaluation bias}\label{sec:reduc-eval-bias}

\subsection{A suggested method to reduce the bias}

A simple transformation of equation (\ref{eq:loss}) shows that for a constant algorithm $g$: $\label{eq:loss:constant} L_t(g) =\sum_{i\in I}P_t(i)l(g_t,i)$. As a consequence, a way to guarantee a stationary evaluation framework for a constant algorithm is to have constant values for the marginal distribution of items, $P_t(i)$.

A natural solution would be to record those probabilities at $t_0$ and use them as the probability to select an item in offline evaluation at $t_1>t_0$. However, as the selection of users and items leads to a joint distribution, this would require to revert the way offline evaluation is done: first select an item, then select a user having this item with a certain probability $\pi_t(u|i)$ leading to a different probability of users selection. Finally this process leads to a similar problem on users, and as in most of systems $\#U > \#I$, it is more efficient to follow the classical evaluation protocol. 

Moreover, we will see that the recalibration of every item is not necessary to reduce the main part of the bias. Indeed in practice most of the time a few items concentrate most of the recommendations (very popular items, discount on selected products, ...). Thus one can reduce the major part of the bias by optimizing the weight of the $p$ items such that the deviation given by $|P_{t_0}(i) - P_{t_1}(i)|$ have the strongest values. In practice $p$ is chosen according to practical constraints (time) or business constraints. 

Thus the weighting strategy that we described in \cite{demytt2014reducing} consists in keeping the classical choice for $P_t(u)$ and weighting $P_t(i|u)$ by departing from the classical values for $P_t(i|u)$ (such as using a uniform probability) in order to mimic static values for $P_ {t_0}(i)$ by :
\begin{equation}
  \label{eq:weighted:conditional}
P_t(i|u,\omega)=\frac{\omega_iP_t(i|u)}{\sum_{j\in I_t}\omega_jP_t(j|u)}.
\end{equation}

These weighted conditional probabilities lead to weighted item probabilities defined by:
\begin{equation}
  \label{eq:weighted:item}
P_t(i|\omega)=  \sum_{u\in  U}P_t(i|u,\omega)P_t(u).
\end{equation}

Then we minimize the distance between $P_{t_1}(i|\omega)$ and $P_{t_0}(i)$ by optimizing the Kullback-Leibler divergence, defined by :
\begin{align}
  \label{eq:divergence}\notag
D(\omega)&=\sum_{i\in I_{t_0}}P_{t_0}(i)\log\frac{P_{t_0}(i)}{P_{t_1}(i|\omega)}
\end{align}
where $I_{t_0}$ represents the set of items present at $t_0$. The asymmetric nature of this distance is useful in our context to consider time $t_0$ as a reference. Moreover this asymmetry reduces the influence of rare items at time $t_0$ (as they were not very important in the calculation of $L_{t_0}(g)$).

\subsection{Previous results}
As described in \cite{demytt2014reducing}, in the classical offline evaluation approach the score of an algorithm in production, given by the classical offline evaluation, tends to increase over time. More generally, the classical offline evaluation  tends to overestimate (resp. underestimate) the unbiased score of an algorithm similar (resp. orthogonal) to the one in production.

We have also shown in \cite{demytt2014reducing} that the suggested weighting strategy perfectly recalibrates the score obtained by the classical offline evaluation for constant algorithms and high values of $p$. Thus, this method seems to reduce the bias for the very simple class of constant algorithms. 

In the next part we apply this method to collaborative filtering algorithms.

\section{Experimentations on a collaborative filtering}\label{sec:exper-eval}

\subsection{Data and metrics}
We consider real world data extracted from Viadeo, where skills are attached to user's profile. The objective of the recommendation systems consists in suggesting new skills to users. The dataset contains 18294 users and 180 items (skills), leading to 117376 couples $(u,i)$. 

Both probabilities $P_t(u)$ and $P_t(i|u)$ are uniform, and the quality function $l$ is given by $l(g_t(u_{-i}),i)=\mathds{1}_{i\in g_t(u_{-i})}$ where $g_t(u_{-i})$ is a set of 5 items. The quality of a recommendation algorithm, $L_t(g)$, is estimated via stochastic sampling in order to simulate what could be done on a larger data set than the one used for this illustration. We selected repeatedly 20 000 couples (user, item) (first we select a user $u$ uniformly, then an item according to $P_t(i|u,\omega)$).

\subsection{Collaborative filtering algorithms}\label{subsec:cf}
Let $X_{u,t}$ be the vector of items of user $u$ at time $t$ ($X_{u,t}  \in \{0,1\}^{\#I}$). Then $X_{u,t}$ is a sparse vector as most of users are associated to only a few items. The objective of collaborative filtering algorithms is to estimate $X_{u,t'}$ for $t'>t$ using the information known on other users. In this paper we will present two different collaborative filtering algorithms:

\[ a) \widehat{X}_{u,t'} = \sum_{v \in U\backslash\{u\}} \frac{\langle X_{u,t}, X_{v,t}\rangle}{\sqrt{\|X_{u,t}\| \cdot \|X_{v,t}\|}}\cdot X_{v,t} \quad  b) \widehat{X}_{u,t'}(i) = \max_{j \in I_u(t)} \frac{\# (U_i \cap U_j)}{\#U_j} \]

The equation $a)$ is known as collaborative filtering with cosine similarity, whereas the equation $b)$ computes the proportion of users associated to item $i$ among the one associated to items possessed by $u$. Then we will note \emph{naive CF} (Collaborative Filtering) the algorithm $b)$.

Finally, the recommendation strategy consists in recommending the $k$ items having the highest values in $\widehat{X}_{u,t'}$.

\subsection{Results}
\begin{figure}[h]
\centering
\subfigure[cosine similarity]{\includegraphics[height=4cm]{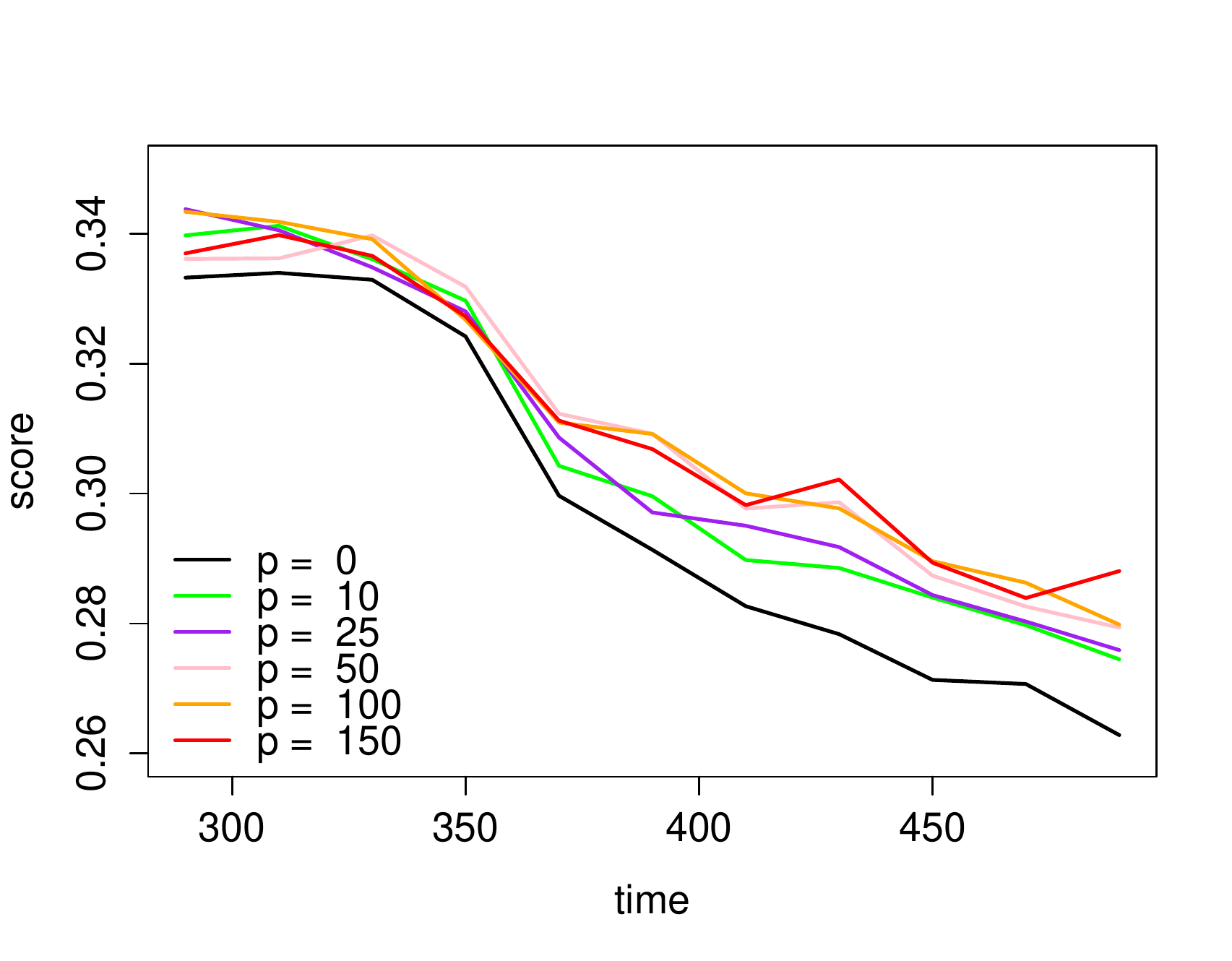}}
\subfigure[naive CF]{\includegraphics[height=4cm]{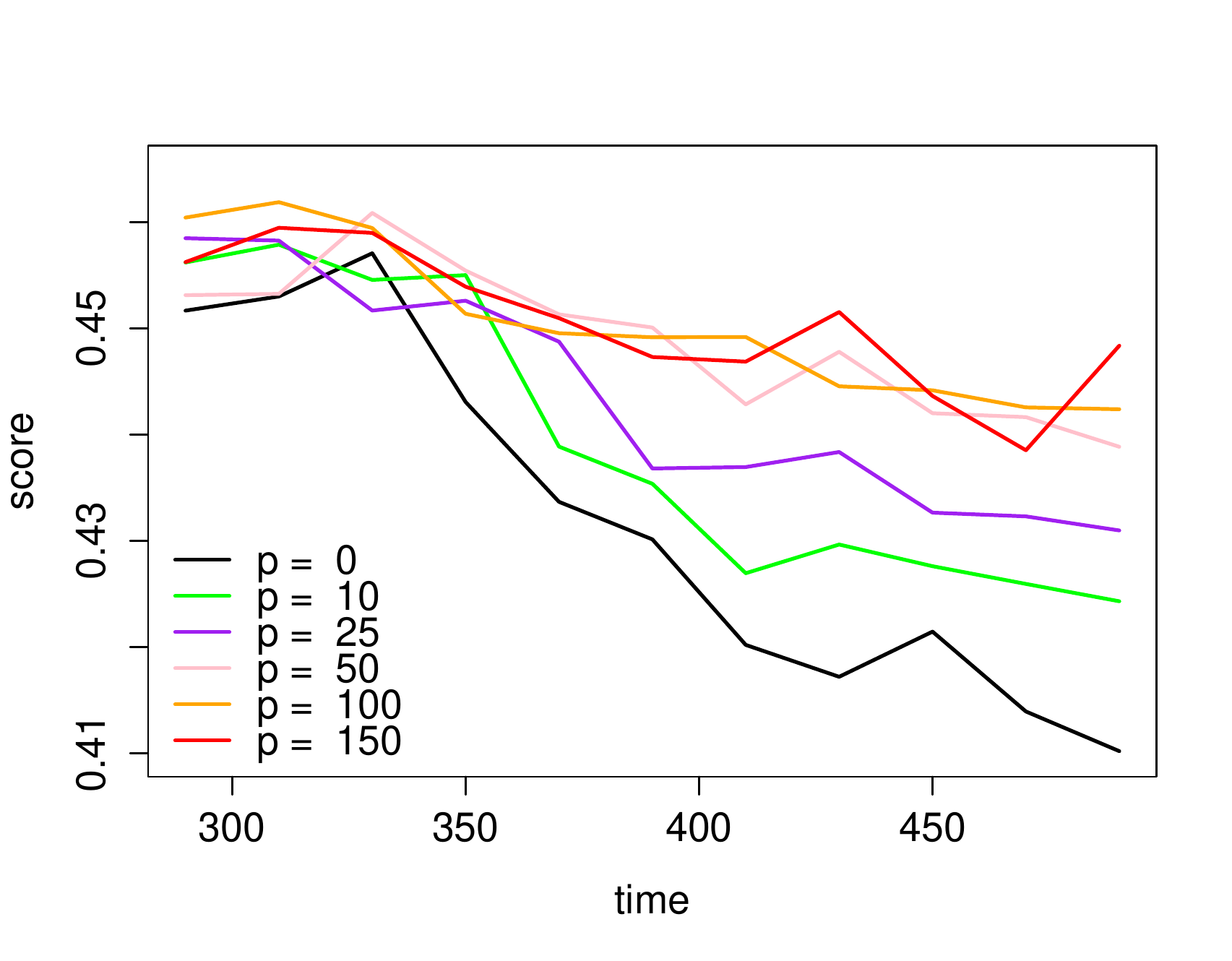}}
\caption{Results on the collaborative filtering with cosine similarity and naive CF, respectively defined by equation $a)$ and $b)$ in section \ref{subsec:cf}, for several values of $p$ (the number of weights optimized).}
\label{fig:cf}
\end{figure}

We apply the method described in Section~\ref{sec:reduc-eval-bias} to compute optimal weights at different instants and for several values of the parameter $p$. The collaborative filtering algorithms are the one presented in section~\ref{subsec:cf}. Results are summarized in figure \ref{fig:cf}. 

The analysis is conducted on a 201 days period, from day 300 to day 500, where day 0 corresponds to the launch date of the skill feature. It is important to notice that two recommendation campaigns were conducted by Viadeo during this period at $t=330$ and $t=430$ respectively. As we can see on figure~\ref{fig:cf}, the scores strongly decrease after the first recommendation campaign ($t = 330$). Thus those campaigns have strongly biased the collected data, leading to a significant bias in the offline evaluation score.

The figure~\ref{fig:cf} shows the influence of the value of $p$: the higher is $p$ the more weights are optimized and the more the bias is corrected. However, the efficiency of the recalibration depends on the algorithms. The results show that the weighting protocol permits to reduce the impact of recommendation campaigns on offline evaluation results as intended. However it does not lead to the stationarity of the score of collaborative filtering algorithms (while it leads to constant scores for constant algorithms). This can be explained by the nature of collaborative filtering: we cannot expect the score to be constant for such an algorithm as it depends on the correlation between users, which have been modified by the recommendation campaigns.

\section{Conclusion}
Various factors influence historical data and bias the score obtained by classical offline evaluation strategy. Indeed, as recommendations influence users, a recommendation algorithm in production tends to be favored by offline evaluation. 

We have presented a new application of the item weighting strategy inspired by techniques designed for tackling the covariate shift problem. Whereas our previous results presented the efficiency of this method for constant algorithms, we have shown that this method also reduces the bias of more elaborate algorithms. 

However the efficiency of this approach depends on algorithms as a recommendation campaign also introduces bias in the correlation between users. Thus the presented strategy reduces a part of the bias, and future works will focus on the structural bias introduced by recommendation campaigns.


\begin{footnotesize}

\bibliographystyle{abbrv}
\bibliography{biblio}

\end{footnotesize}


\end{document}